\documentclass[superscriptaddress,twocolumn,amsmath,amssymb]{revtex4}
\usepackage{graphicx}
\usepackage{color}

\renewcommand{\vec}[1]{\mbox{\boldmath $#1$}}

\begin{document}

\title{New concept for pairing anti-halo effect as a localized 
wave packet of quasi-particle} 

\author{K. Hagino}
\affiliation{ 
Department of Physics, Tohoku University, Sendai, 980-8578,  Japan} 
\affiliation{Research Center for Electron Photon Science, Tohoku University, 1-2-1 Mikamine, Sendai 982-0826, Japan}
\affiliation{
National Astronomical Observatory of Japan, 2-21-1 Osawa,
Mitaka, Tokyo 181-8588, Japan}

\author{H. Sagawa}
\affiliation{
RIKEN Nishina Center, Wako 351-0198, Japan}
\affiliation{
Center for Mathematics and Physics,  University of Aizu, 
Aizu-Wakamatsu, Fukushima 965-8560,  Japan}


\begin{abstract}
The pairing anti-halo effect is a phenomenon that a pairing correlation 
suppresses a divergence of nuclear radius, which happens 
for single-particle states with orbital angular momenta of $l$ = 0 and 1
in the limit of 
vanishing binding energy. 
While this effect has mainly been discussed in terms of 
Hartree-Fock-Bogoliubov (HFB) theory, we here 
use a three-body model and 
provide 
its new  intuitive  concept as a localized wave packet for a 
quasi-particle, that is, 
a coherent superposition of a weakly bound 
and continuum 
wave functions due to a pairing interaction. 
We show that 
the one-particle density in the three-body model 
can be directly expressed with such 
quasi-particle wave functions, 
which have a close 
analog to wave functions in the HFB approximation. 
\end{abstract}


\maketitle

\section{Introduction}

It has been known well that, 
in the limit of vanishing 
binding energy, 
the root-mean-square radius diverges 
for a wave function in a short range potential well with orbital angular 
momenta of $l$ = 0 and 1 \cite{RJM92,Jensen04}. 
Halo nuclei, which are characterized 
by an extended density distribution \cite{Tanihata85,Tanihata88}, 
have been interpreted to be due to such divergence with a single-particle 
wave function for 
$l$ = 0 and 1 \cite{S92}. See Refs. \cite{HTS13,TSK13,SH15} for 
recent review articles on halo nuclei. 

For even-even nuclei, the pairing correlation among valence neutrons 
plays a decisive role in 
the structure of weakly bound nuclei \cite{Doba96,BE91,EBH97,HS05}. 
Bennaceur, Dobaczewski, and Ploszajczak have demonstrated 
that 
the root-mean-square radius does not diverge for even-even nuclei as 
the pairing correlation largely suppresses the halo structure 
in odd-mass nuclei,  
which has been referred to as 
the pairing anti-halo effect \cite{BDP00}. 
In Refs. \cite{HS11,HS12,HS12-2}, 
we have argued that the odd-even 
staggerings observed in reaction cross sections \cite{Takechi12,Takechi14} 
can be interpreted in terms of the pairing anti-halo 
effect (see also Refs. \cite{Sasabe13,MY14}). 

The pairing anti-halo effect has been studied 
using mainly the 
Hartree-Fock-Bogoliubov (HFB) method \cite{BDP00,CRM14,GYSG06,Y05}. 
In this approach, the pairing anti-halo effect occurs because the 
quasi-particle energy remains a finite value 
even when a single-particle energy vanishes\cite{BDP00}. 
A key issue for this argument is that the pairing gap needs to be finite 
in the zero binding limit. Many HFB calculations have actually shown 
that it is indeed the case \cite{BDP00,HS11,HS12,HS12-2,CRM14,GYSG06,Y05}, 
leading to a suppression of the halo structure in even-even systems. 

Although the HFB method provides a clear mathematical interpretation 
of the pairing anti-halo effect, its physical mechanism 
is 
less transparent.  The aim of this paper is to propose a more intuitive idea 
on the pairing anti-halo effect, using a three-body model 
with a core nucleus and two valence neutrons. This model is formulated 
with a simpler single-particle basis, providing a complementary 
interpretation to the one based on the HFB method. 

The paper is organized as follows. 
In Sec. II, we first show how the pairing anti-halo effect is realized in 
the HFB method. We then introduce the three-body model and discuss 
the pairing anti-halo effect in this model. 
In Sec. III, we introduce a quasi-particle 
wave function within the three-body model and investigate its structure. 
We show that a coherent superposition of a weakly bound state and 
continuum states is a key ingredient of the pairing anti-halo effect. 
We then summarize the paper in Sec. IV. 

\section{Pairing anti-halo effect}

\subsection{Hartree-Fock-Bogoliubov method}

Before we discuss the pairing anti-halo effect with a three-body model, we 
first show how it is understood in the HFB method. This is also to clarify 
the notation used in this paper. 

For a two-body system 
which consists of a valence neutron and a core nucleus, 
we assume that the wave function for the relative motion obeys the 
Schr\"odinger equation given by, 
\begin{equation}
\hat{h}\,\psi_{nljm}(\vec{r}) = \left[-\frac{\hbar^2}{2\mu}\vec{\nabla}^2+V(r)
\right]\psi_{nljm}(\vec{r})
=\epsilon_{nlj}\,\psi_{nljm}(\vec{r}), 
\label{two-body}
\end{equation}
where $\mu$ is the reduced mass and $V(r)$ is the potential between the 
valence neutron and the core nucleus. We have assumed that $V(r)$ is local 
and has 
spherical symmetry so that the wave function is characterized by 
the orbital angular momentum $l$, the total angular momentum $j$ and 
its $z$-component, $m$, as well as the radial quantum number, $n$. 
Here, $\epsilon_{nlj}$ is the energy eigenvalue. 

For simplicity, we consider only an $s$ wave solution of this 
Schr\"odinger equation. 
The radial wave function $u_{nlj}(r)$, defined 
with the spin-angular function, ${\cal Y}_{jlm}(\hat{\vec{r}})$, by 
\begin{equation}
\psi_{nljm}(\vec{r})=\frac{u_{nlj}(r)}{r}{\cal Y}_{jlm}(\hat{\vec{r}}), 
\label{radialwf}
\end{equation}
 behaves asymptotically as, 
\begin{equation}
u_{nlj}(r)\sim \exp(-\alpha_{nlj}\,r),
\label{radialwf2}
\end{equation}
where $\alpha_{nlj}$ is defined as 
$\alpha_{nlj}=\sqrt{2\mu|\epsilon_{nlj}|/\hbar^2}$. 
The expectation value of $r^2$ then reads,
\begin{equation}
\langle r^2\rangle  \sim \frac{1}{2\alpha_{nlj}^2}=
\frac{\hbar^2}
{4\mu|\epsilon_{nlj}|}, 
\end{equation}
which apparently diverges in the limit of $\epsilon_{nlj}\to 0$. 

In many-body systems with the pairing correlation, one may consider the 
HFB equations given by \cite{Doba96,Doba84}, 
\begin{equation}
\left(
\begin{array}{cc}
\hat{h}-\lambda&\Delta(r) \\
\Delta(r)&-\hat{h}+\lambda
\end{array}
\right)
\left(
\begin{array}{c}
U_{nljm}(\vec{r})\\
V_{nljm}(\vec{r})
\end{array}
\right)
=E_{nlj}
\left(
\begin{array}{c}
U_{nljm}(\vec{r})\\
V_{nljm}(\vec{r})
\end{array}
\right),
\label{HFB}
\end{equation}
where $\hat{h}$ is the mean-field Hamiltonian given in Eq. (\ref{two-body}), 
$\Delta(r)$ is the 
pairing potential, and $\lambda$ is the chemical potential. 
We have again assumed that both the mean-field potential in $\hat{h}$ and 
the pairing potential, $\Delta(r)$, are spherical and local functions 
of $r$. 
In the HFB equations, Eq. (\ref{HFB}), $E_{nlj}$ is a quasi-particle energy 
and $U_{nljm}(\vec{r})$ and $V_{nljm}(\vec{r})$ are the upper and the lower 
components of a quasi-particle wave function, respectively. 
These are ortho-normalized according to, 
\begin{equation}
\int d\vec{r}[U^*_\alpha(\vec{r})U_\beta(\vec{r}) 
+V^*_\alpha(\vec{r})V_\beta(\vec{r})] =\delta_{\alpha,\beta},
\label{normalizationHFB}
\end{equation}
where $\alpha$ and $\beta$ are shorthanded notations for $(njlm)$. 
The one-particle density, $\rho(r)$, is given in terms of $V_{nljm}(\vec{r})$ 
by, 
\begin{equation}
\rho(r)=\sum_n\sum_{j,l,m}|V_{nljm}(\vec{r})|^2.
\label{rho-hfb}
\end{equation}

In the BCS approximation, 
$V_{nljm}(\vec{r})$ is expressed by a product of the occupation 
factor $v_{nlj}^{BCS}$ and the single-particle wave 
function, $\psi_{nljm}(\vec{r})$ \cite{Doba96,HS05-2}. 
In contrast, in the HFB, 
the asymptotic form of the radial part for the lower 
component, defined similarly to Eq. (\ref{radialwf2}), reads \cite{Doba96}, 
\begin{equation}
v_{nlj}(r)\sim \exp(-\beta_{nlj}\,r),
\label{low-component}
\end{equation}
for $l=0$, where $\beta_{nlj}$ is given as 
$\beta_{nlj}=\sqrt{2\mu(E_{nlj}-\lambda)/\hbar^2}$. 
The expectation value of $r^2$ with this wave function  reads, 
\begin{equation}
\langle r^2\rangle  \sim \frac{1}{2\beta_{nlj}^2}. 
\end{equation}
Notice that the quasi-particle energy, $E_{nlj}$, is given in the BCS 
approximation as, $E_{nlj}=\sqrt{(\epsilon_{nlj}-\lambda)^2+\Delta_{nlj}^2}$, 
where $\Delta_{nlj}$ is the pairing gap. This implies that $E_{nlj}$ behaves 
as $E_{nlj}\sim\Delta_{nlj}$ in the zero binding limit with 
$\lambda \sim \epsilon_{nlj} \sim 0$. The expectation value of $r^2$ then 
reads, 
\begin{equation}
\langle r^2\rangle  \sim 
\frac{\hbar^2}{4\mu\,\Delta_{nlj}}, 
\end{equation}
which remains finite as long as the pairing gap, $\Delta_{nlj}$, is finite. 
This is nothing but the pairing anti-halo effect proposed 
in Ref. \cite{BDP00}. An essential point for the pairing anti-halo effect is 
that the single-particle energy, $\epsilon_{nlj}$, is replaced by the 
quasi-particle energy, $E_{nlj}-\lambda$, reflecting the pairing 
correlation, which then induces a shrinkage of wave function 
according to Eq. (\ref{low-component}). 

\subsection{Three-body model}

In order to achieve a simple but still physical  concept 
for the 
pairing anti-halo effect, let us 
now introduce a three-body model which consists of the core nucleus and 
two valence neutrons. 
The Hamiltonian for the three-body model reads \cite{EBH97,HS05},
\begin{equation}
H=\hat{h}(1)+\hat{h}(2)+v_{\rm pair}(\vec{r}_1,\vec{r}_2)+ 
\frac{\vec{p}_1\cdot\vec{p}_2}{m_c},
\end{equation}
where the single-particle Hamiltonian, $\hat{h}$, is the same as the one 
in Eq. (\ref{two-body}) and 
$v_{\rm pair}(\vec{r}_1,\vec{r}_2)$ is a pairing interaction between the 
two valence neutrons. The last term in this equation is the two-body part 
of the recoil kinetic energy of the core nucleus, whose mass is 
denoted by $m_c$. 

Using the eigen-functions of the single-particle Hamiltonian $\hat{h}$, 
that is, 
the wave functions $\psi_{nljm}(\vec{r})$ in Eq. (\ref{two-body}), 
the two-particle wave function for the ground state of the three-body 
system with spin-parity of $J^\pi=0^+$ is given as, 
\begin{equation}
\Psi(\vec{r}_1,\vec{r}_2) = \sum_{n,n',l,j} C_{nn'lj}
\Psi_{nn'lj}^{(2)}(\vec{r}_1,\vec{r}_2), 
\label{twoparticle}
\end{equation}
with
\begin{eqnarray}
\Psi_{nn'lj}^{(2)}(\vec{r}_1,\vec{r}_2)&=&
[\psi_{nlj}(\vec{r}_1)\psi_{n'lj}(\vec{r}_2)]^{J=0}, \\
&=&\sum_m\frac{(-1)^{j-m}}{\sqrt{2j+1}}\,
\psi_{nljm}(\vec{r}_1)\psi_{n'lj-m}(\vec{r}_2). \nonumber \\
\end{eqnarray}
(For simplicity of the notation, we do not use here the anti-symmetrized 
basis \cite{BE91}. The anti-symmetrization is realized by setting 
$C_{nn'jl}=C_{n'nlj}\equiv \tilde{C}_{nn'lj}/\sqrt{2}$ in Eq. (\ref{twoparticle})). 
The one-particle density constructed with this two-particle 
wave function is then given by \cite{BE91}, 
\begin{eqnarray}
\rho(\vec{r})&=&\int d\vec{r}'|\Psi(\vec{r},\vec{r}')|^2, \\
&=& 
\sum_{n,n',\tilde{n}}\sum_{j,l,m}\frac{C^*_{nn'lj}C_{\tilde{n}n'lj}}{2j+1}\,
\psi_{nljm}^*(\vec{r})\psi_{\tilde{n}ljm}(\vec{r}). \nonumber \\
\label{rho1-0}
\end{eqnarray}
Using the spherical reduction for the wave functions 
(see Eq. (\ref{radialwf})), one can show that the one-particle density 
is expressed as \cite{BE91}, 
\begin{equation}
\rho(r)=
\frac{1}{4\pi}\,\sum_{n,n',\tilde{n}}\sum_{j,l,m}C^*_{nn'lj}C_{\tilde{n}n'lj} 
\phi^*_{nlj}(r)\phi_{\tilde{n}lj}(r),
\label{rho1}
\end{equation}
where $\phi_{nlj}$ is defined as $\phi_{nlj}(r)=u_{nlj}(r)/r$. 

\begin{figure} 
\includegraphics[scale=0.6,clip]{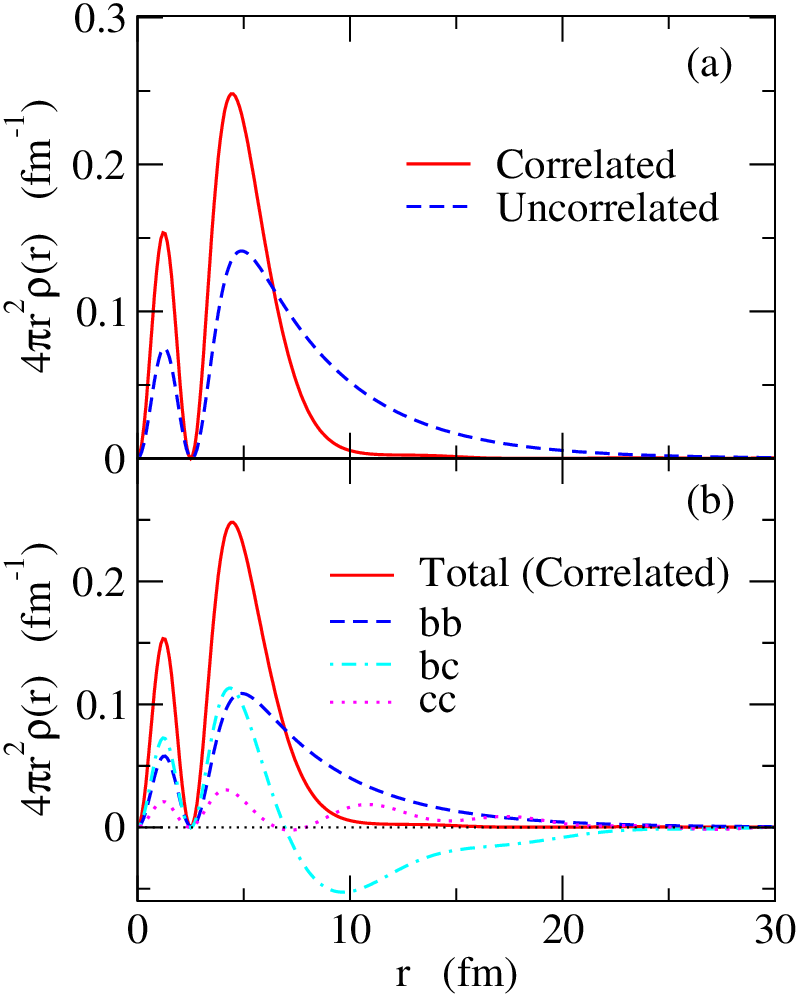}
\caption{
(Upper panel) 
The correlated (the solid line) and the uncorrelated (the dashed line) 
one-particle densities obtained with the three-body model for $^{24}$O. 
Only the $s$-wave single-particle states are included in the calculations. 
In the uncorrelated case, the two valence 
neutrons occupy the 2$s_{1/2}$ state 
at $\epsilon = -0.275$ MeV, while they are scattered into the continuum 
states in the correlated case. A zero-range pairing interaction is employed, which 
yields the ground state energy of $E_{\rm g.s.} = -2.46$ MeV. 
(Lower panel) 
The decomposition of the correlated density into three components. 
The (bb) component (the dashed line) 
corresponds to the one in which both of the 
two valence neutrons occupy the bound 2$s_{1/2}$ state, while in the (bc) 
component shown by the dot-dashed line, 
one of them is scattered to a continuum state. The (cc) 
component shown by the dotted line 
corresponds to the one in which both of the valence neutrons 
are scattered into continuum states. The total correlated density 
is also shown by the solid line. }
\end{figure}

The upper panel of Fig. 1 shows a one-particle density in the three-body 
model. To draw this figure, we consider the $^{24}$O nucleus 
($^{24}$O = $^{22}$O + $n$ + $n$), 
and employ a contact pairing interaction, 
$v_{\rm pair}(\vec{r},\vec{r}')=-g\delta(\vec{r}-\vec{r}')$, with 
$g = 1374$ MeV fm$^3$, together with the cutoff 
energy of $E_{\rm cut} = 10$ MeV. The continuum states are discretized with 
the box boundary condition with the box size of $R_{\rm box}$ = 30 fm. 
We use a Woods-Saxon potential for 
the mean-field potential, $V(r)$, with the radius parameter of 
$R$ = 3.5 fm and the diffuseness parameter of $a$ = 0.67 fm \cite{HS05}. 
The depth of the Woods-Saxon potential is somewhat arbitrarily chosen to 
be $V_0$ = $-34.56$ MeV, which has the 2$s_{1/2}$ state at $-0.275$ MeV. 
For simplicity of the discussion, we include only $l$ = 0 in 
Eq. (\ref{twoparticle}), for which the 1$s_{1/2}$ state is 
assumed to be occupied by the core nucleus and is explicitly excluded in 
the summation. In this model space, there are
 one bound state, 2$s_{1/2}$, 
at $\epsilon=-0.275$ MeV, and five discretized $s$-wave continuum states 
up to 10 MeV. This calculation yields the ground state energy 
of $E_{\rm g.s.} = -2.46$ MeV. 

The dashed line in the figure shows the one-particle density in the 
absence of the pairing interaction, which is proportional to the square 
of the 2$s_{1/2}$ wave function, that is, 
$\rho(r)=|\phi_{2s_{1/2}}(r)|^2/4\pi$. Since the 2$s_{1/2}$ state is a weakly 
bound $s$-wave state, the resultant density has an extended long tail. 
In contrast, in 
the correlated density distribution shown by the solid line, the density 
distribution is considerably shrunk compared to the uncorrelated density. 
The root-mean-square radii are $\sqrt{\langle r^2\rangle}$ = 5.18 and 8.83 fm 
for the correlated and the uncorrelated cases, respectively. 
This is a clear manifestation of 
the pairing anti-halo effect discussed in the previous 
sub-section. 

The lower panel of Fig. 1 shows a decomposition of the correlated 
density. Here, we decompose it into three components, that is, 
(i) (bb): both $n$ and $\tilde{n}$ in Eq. (\ref{rho1}) 
belong to the weakly bound state, 2$s_{1/2}$, (ii) (bc): one of them belongs to the bound state while 
the other belongs to a continuum state, and (iii) (cc): both of them 
belong to continuum states. 
The (bb) component in fact has the same radial 
profile as  the uncorrelated density 
shown in the upper panel, having an extended tail. The (bc) component 
behaves similarly to the (bb) component inside the potential, while it has 
the opposite sign to the (bb) component in the tail region. Because of 
this, the density in the inner part is enhanced while the (bb) and (bc) components 
are largely canceled out in the outer 
part. 
One can thus find that the scattering of a particle to the continuum 
spectrum due to the pairing interaction plays an essential role in the 
pairing anti-halo effect. The (cc) component, on the other hand, provides 
only a small portion of the correlated density, even though it is not 
negligible. This component is positive in a wide range of radial 
coordinate, as one can see in the figure. 

\section{Quasi-particle wave function in the three-body model}

In order to get a deeper insight into the pairing anti-halo 
effect in the three-body model, we next re-express the one-particle 
density in a different form. To this end, we first 
notice that the two-particle 
wave function, Eq. (\ref{twoparticle}), is expressed as, 
\begin{equation}
\Psi(\vec{r}_1,\vec{r}_2) = \sum_{n'}\sum_{l,j} \,
[\tilde{\psi}_{n'lj}(\vec{r}_1)\psi_{n'lj}(\vec{r}_2)]^{J=0}, 
\end{equation}
with 
\begin{equation}
\tilde{\psi}_{n'ljm}(\vec{r})\equiv
\sum_n C_{nn'lj}\,\psi_{nljm}(\vec{r}). 
\end{equation}
The one-particle density, Eqs. (\ref{rho1-0}) and (\ref{rho1}), 
is then given as, 
\begin{eqnarray}
\rho(\vec{r})&=&
\sum_{k}\sum_{j,l,m}\frac{1}{2j+1}\,
|\tilde{\psi}_{kljm}(\vec{r})|^2, \\
&=&\frac{1}{4\pi}\,\sum_{k}\sum_{j,l}
\left|\frac{\tilde{u}_{klj}(r)}{r}\right|^2, 
\end{eqnarray}
where 
$\tilde{u}_{klj}(r)$ is defined as, 
\begin{equation}
\tilde{u}_{klj}(r)\equiv
\sum_n C_{nklj}\,u_{nlj}(r). 
\label{qpwf-radial}
\end{equation}
Notice that this is in a similar form to the one-particle wave function 
in the Hartree-Fock-Bogoliubov approximation, especially if one 
expands the quasi-particle wave function, $V_{kljm}$, on the Hartree-Fock 
basis, $\psi_{nljm}$ \cite{Gall94,Terasaki96,HS05-2}. 
For this reason, we shall call 
$\tilde{\psi}_{n'ljm}(\vec{r})$ a ``quasi-particle'' 
wave function hereafter. 
Notice that the quasi-particle wave functions 
$\tilde{\psi}_{nljm}$ are not orthonormalized,  
just the same as  the HFB wave functions 
$V_{nljm}$ (see Eq. (\ref{normalizationHFB})). 

\begin{figure} 
\includegraphics[scale=0.65,clip]{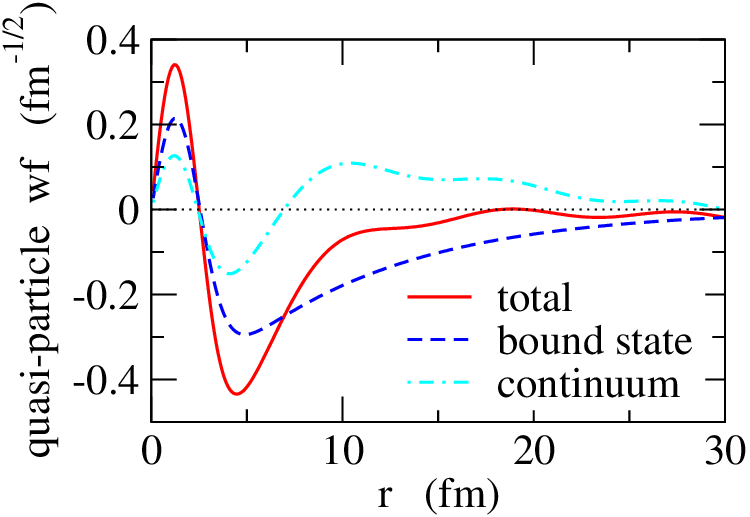}
\caption{
The radial part of the quasi-particle wave function, $\tilde{u}_{2s_{1/2}}(r)$, 
defined 
by Eq. (\ref{qpwf-radial}), for the weakly bound 2$s_{1/2}$ state. The solid 
line shows the total wave function, while the dashed and the dot-dashed 
lines denote its bound state and continuum state contributions 
as defined by Eq. (\ref{decomposition}), 
respectively. 
}
\end{figure}

The solid line in Fig. 2 shows the radial dependence of the quasi-particle 
wave function for the weakly-bound 2$s_{1/2}$ state, that is, 
$\tilde{u}_{klj}(r)$ with $(klj)=2s_{1/2}$, 
for the three-body Hamiltonian introduced in the previous subsection. 
The dashed and the dot-dashed lines show its decomposition into the bound 
state and the continuum state contributions, respectively. They are 
defined as 
\begin{eqnarray}
\tilde{u}_{klj}(r)&=&
\tilde{u}^{(b)}_{klj}(r)+\tilde{u}^{(c)}_{klj}(r), 
\label{decomposition}
\\
&=&\sum_{n=2s_{1/2}} C_{nklj}\,u_{nlj}(r)+\sum_{n={\rm cont.}} 
C_{nklj}\,u_{nlj}(r). \nonumber \\
\end{eqnarray}
One can see that the main feature of this quasi-particle wave function 
is similar to the one-particle density shown in Fig. 1 (b). That is, 
the bound state and the continuum state 
contributions are largely canceled with 
each other outside the potential while the two components 
contribute coherently in the inner region. 
We notice that the localization due to a coherent superposition 
of continuum states 
is the same mechanism as a formation of a localized wave packet. 
This is an essential ingredient of the pairing anti-halo 
effect, that is, a formation of localized wave packet 
induced by 
a pairing interaction. 

\begin{figure} 
\includegraphics[scale=0.6,clip]{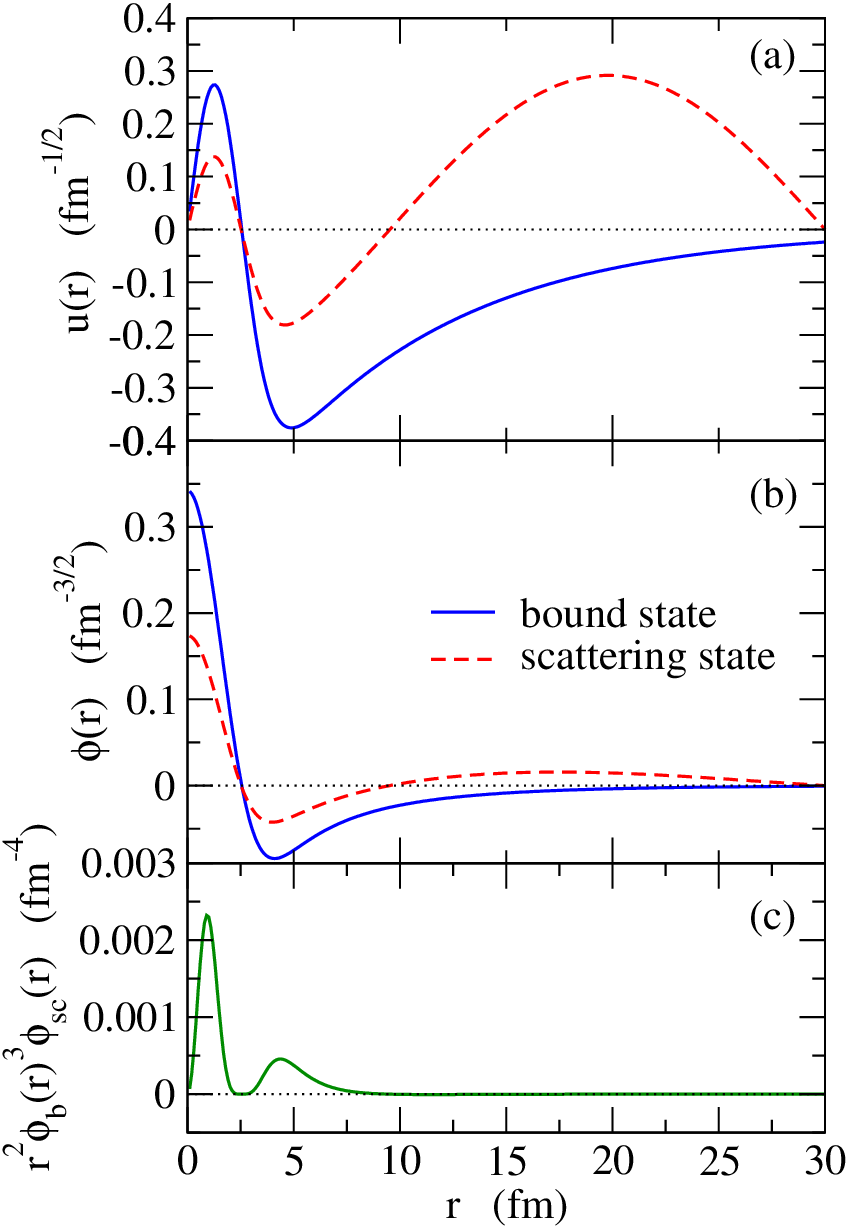}
\caption{
Figs. 3(a) and 3(b): The radial component of the 
wave functions for the weakly bound 
2$s_{1/2}$ state (the solid line) and the lowest discretized $s$-wave 
state at $\epsilon$ = 0.51 MeV (the dashed line). Fig. 3(b) shows the 
radial wave function $\phi(r)$ while Fig. 3(a) shows $u(r)\equiv r\phi(r)$. 
Fig. 3(c): the integrand of the matrix element for the zero-range pairing 
interaction between the components $[\psi_b(\vec{r})\psi_b(\vec{r}')]$ and 
${\cal A}[\psi_b(\vec{r})\psi_c(\vec{r}')]$, where ${\cal A}$ is the anti-symmetrizer 
and $\psi_b$ and $\psi_c$ are the wave functions for the weakly bound state and the 
lowest continuum state, respectively. 
}
\end{figure}

A question still remains concerning why 
the superposition is in such a way that the tail part of the bound 
wave function is suppressed. In order to clarify this point, let us 
strict ourselves only to 
two single-particle states, one is the weakly 
bound 2$s_{1/2}$ state at $\epsilon_b=-0.275$ MeV 
and the other is the lowest discretized 
$s$-wave continuum state. For the potential given in the previous section, 
the latter state is at $\epsilon_c$ = 0.51 MeV for $R_{\rm box}$ = 30 fm. 
Figs. 3(a) and 3(b) show the radial part of the wave functions 
for these states. The solid and the dashed lines correspond to the bound 
state and the scattering states, respectively. Fig. 3(b) shows the radial 
wave functions, $\phi(r)$, while Fig. 3(a) shows $u(r)=r\phi(r)$. 
In the inner part, the two wave functions behave similarly to each other, 
because the absolute value of the single-particle energies, $|\epsilon|$, 
is small for both the states, so that $V(r)-\epsilon\sim V(r)$. 
In the outer region where the potential $V(r)$ disappears, the two 
wave functions should behave differently. 
Since they behave similarly in the inner region, 
the two wave functions have to have opposite 
sign in the outer region in order to fulfill the orthogonal condition. 

With these two single-particle states, 
we assume, for simplicity, that the two-particle wave function is 
given by,
\begin{equation}
\Psi(\vec{r}_1,\vec{r}_2)=C_{bb}\,\psi_b(\vec{r}_1)\psi_b(\vec{r}_2)
+C_{bc}\,{\cal A}[\psi_b(\vec{r}_1)\psi_c(\vec{r}_2)], 
\label{twoparticle-2}
\end{equation}
where $\psi_b$ and $\psi_c$ are the bound and the scattering wave 
functions, respectively, and ${\cal A}$ is the anti-symmetrizer. 
The coefficients $C_{bb}$ and $C_{bc}$ are obtained by solving the 
eigen-value equation, 
\begin{equation}
\left(
\begin{array}{cc}
\epsilon_{bb}& F \\
F& \epsilon_{bc}
\end{array}
\right)
\left(
\begin{array}{c}
C_{bb}\\
C_{bc}
\end{array}
\right)
=E 
\left(
\begin{array}{c}
C_{bb}\\
C_{bc}
\end{array}
\right),
\label{2x2}
\end{equation}
where $\epsilon_{bb}$ and $\epsilon_{bc}$ are the diagonal components 
of the three-body Hamiltonian 
including the pairing matrix elements (in the 
present case, $\epsilon_{bb}$ and $\epsilon_{bc}$ are $-2.16$ and $-0.53$ 
MeV, respectively). 
$F$ is the matrix element of the pairing interaction 
between the two configurations, that is, 
\begin{equation}
F=-\frac{g}{4\pi}\int^\infty_0r^2dr\,[\phi_b(r)\phi_b(r)]^*[\phi_b(r)\phi_c(r)]. 
\end{equation}
The integrand is shown in Fig. 3(c). As one can see, the integrand is 
positive, except for large values of $r$, for which the contribution 
is negligibly small, and thus $F$ is negative for an attractive pairing 
interaction with $g >$  0. 

The eigen-values $E$ and the corresponding eigen-vectors of 
Eq. (\ref{2x2}) read, 
\begin{eqnarray}
E_\pm &=& 
\frac{1}{2}\,\left\{(\epsilon_{bb}+\epsilon_{bc})
\pm\sqrt{(\epsilon_{bc}-\epsilon_{bb})^2+4F^2}\right\}, \nonumber \\
\\
\left(
\begin{array}{c}
C_{bb}\\
C_{bc}
\end{array}
\right)
&=&{\cal N}
\left(
\begin{array}{c}
F\\
E_\pm-\epsilon_{bb}
\end{array}
\right),
\label{eigenvector}
\end{eqnarray}
where ${\cal N}$ is the normalization factor. For the lower eigen-value, 
$E_-$, the quantity $E_--\epsilon_{bb}$ reads
\begin{equation}
E_--\epsilon_{bb} 
=
\frac{1}{2}\,\left\{(\epsilon_{bc}-\epsilon_{bb})
-\sqrt{(\epsilon_{bc}-\epsilon_{bb})^2+4F^2}\right\}, 
\end{equation}
which is apparently negative. 
Since $F$ and $E_--\epsilon_{bb}$ are both negative, 
$C_{bb}$ and $C_{bc}$ thus have the same 
sign to each other (see Eq. (\ref{eigenvector})), 
leading to the quasi-particle wave function which has a suppressed 
tail as shown in Fig. 2. This feature 
remains the case even when higher continuum states and/or the 
configuration with $\psi_c(\vec{r}_1)\psi_c(\vec{r}_2)$ are included in the two-particle wave function. 

There is a freedom for the phase of single-particle 
wave functions to take a positive value or a negative value
at the origin.  
In Fig. 3(b), the two $s_{1/2}$ wave functions are taken to be 
positive at the origin.
We notice that 
the shrinkage of the halo wave function 
is independent of the choice of the sign of the wave function. 
That is, 
if one takes the negative sign 
for the
continuum wave function at the origin,  
the sign of the pairing matrix $F$  turns to be positive so that 
$C_{bb}$ and $C_{bc}$ have a different sign from one another. 
However, the one particle density as well as the quasi-particle 
wave function remain the same, 
since 
the sign of the amplitude $C_{bb}$ in Eq. (\ref{eigenvector}) 
and that of the single-particle wave function, $\psi_c(\vec{r}_2)$, 
in Eq. (\ref{twoparticle-2}) are simultaneously altered, whereas 
the sign of $C_{bc}$ remains the same.

\section{Summary}

We have discussed the pairing anti-halo effect 
from a three-body model perspective. 
In contrast to the conventional understanding 
based on a Hartree-Fock-Bogoliubov (HFB)
wave function, the present study  provides a simple and 
intuitive concept  for the pairing anti-halo effect.  
Namely, we have found that an essential 
ingredient of the pairing anti-halo effect is a coherent 
superposition of a loosely-bound and 
continuum states due to a pairing interaction, which 
leads to a localized wave function as a wave packet. 
The coherence of the wave functions results in an enhancement of 
one-particle density in the inner region while the long tail of a weakly bound wave function is 
largely canceled out with continuum wave functions. 
The present study offers a complementary understanding 
for the 
pairing anti-halo effect to the one 
with the HFB approximation.   
In fact, 
we have shown that the one-particle density 
with a three-body model can be cast into a similar 
form of the density in the Hartree-Fock-Bogoliubov 
approximation. We have pointed out that such 
``quasi-particle'' wave functions show the shrinkage effect
as a consequence of 
a coherent superposition 
of a weakly bound  and continuum states. 

\section*{Acknowledgments}

This work was partly supported by   by JSPS KAKENHI  Grant Numbers  JP16K05367.

\end{document}